\documentclass[12pt]{article}

\usepackage{latexsym}

\usepackage{graphics}
\usepackage{graphicx}
\usepackage{epstopdf}

\begin{document}


\title{On the Thermodynamics of 5D Black Holes on ALF Gravitational Instantons}

\author{
    Petya G. Nedkova\thanks{E-mail:pnedkova@phys.uni-sofia.bg}, Stoytcho S. Yazadjiev \thanks{E-mail: yazad@phys.uni-sofia.bg}\\
{\footnotesize  Department of Theoretical Physics,
                Faculty of Physics, Sofia University,}\\
{\footnotesize  5 James Bourchier Boulevard, Sofia~1164, Bulgaria }\\
}

\date{}

\maketitle

\begin{abstract}
We investigate the thermodynamical properties of solutions to the 5D
Einstein and Einstein-Maxwell-dilaton equations describing  static black
holes on ALF gravitational instantons. Some general expressions
regarding the nut charge and potential are derived, as well as
Smarr-like relations. The thermodynamics of two particular solutions
representing a static black hole on Taub-Nut and Taub-Bolt
instantons are examined in detail. Two different approaches for computation of the mass and the tension - the Komar and counterterm method, are applied and compared. A new charged black hole on
Taub-Bolt instanton is constructed and investigated.
\end{abstract}


\sloppy

\section{Introduction}

\paragraph{}In recent years higher dimensional gravity has attracted a lot of interest and research endeavors.
 In contrast to the 4-dimensional case, the variety of possible self-gravitating
systems is much more abundant. Curious systems of black holes with
diverse topology have been discovered
 \cite{Emparan:2008}, as well as
systems of black holes held apart or 'sitting on' vacuum objects
called spacetime bubbles  \cite{Elvang:2002}-\cite{Nedkova:2010}.
Mathematically, the spacetime bubbles are the simplest case of
gravitational instantons, known in 4D as Euclidian Schwarzchild
instantons. Research further expanded into seeking solutions to the
higher dimensional Einstein-Maxwell-dilaton equations which describe systems
of black holes and more complex gravitational instantons. Using
various solution generation methods, the so called squashed black
holes and their charged counterparts were obtained
\cite{Ishihara:2005}, \cite{Wang:2006}, \cite{Yazadjiev:2006}, \cite{Stelea:2011},
representing  black holes on Taub-Nut instanton background, as well
as black holes on Kerr, Eguchi-Hanson, Taub-Bolt and
multi-instantons \cite{Chen:2011}, \cite{Ishihara:2006}. A recent
description of the 5D vacuum solutions describing black holes on
gravitational instantons was made by Chen and Teo in their work
\cite{Chen:2011}.

The instanton solutions are characterized with more involved geometry leading to interesting physical properties.
They possess complicated asymptotics, the spacetime being twisted in a nontrivial way, which makes their thermodynamics an interesting area of research.
The goal of the current article is to examine the thermodynamical properties of a class of solutions to the 5D Einstein and Einstein-Maxwell-dilaton equations representing a static black hole on a asymptotically locally flat gravitational instanton.

The article is organized as follows. In the first sections we introduce some basic properties of the gravitational instantons and black holes on gravitational
instantons which are crucial for the investigation of their thermodynamics. Next,  we discuss the relevant thermodynamical characteristics and derive some useful
expressions for the nut charge and potential, as well as a Smarr-like formula. In section 5 we apply the general results of the previous
section to examine the properties of two particular vacuum solutions describing a static black hole on a Taub- Nut, or Taub-Bolt instanton.
The final section is devoted to the construction of a new solution describing a charged black hole on a Taub-Bolt instanton, and investigation
of its thermodynamics.

\section{Classification of gravitational instantons}

\paragraph{}Gravitational instantons are defined as non-singular Euclidean solutions to the 4-dimensional Einstein equations possessing a finite action.
There exist different classifications of gravitational instantons. The first one was made by Gibbons and Hawking \cite{Gibbons:1979c} by means of their symmetries.
Nearly all the known gravitational instantons possess a $U(1)\times U(1)$ symmetry group generated by a couple of commuting spacelike Killing fields. Gibbons
and Hawking, however, chose to build a classification scheme based only on the action of a 1-parameter isometry group, or more precisely on its fixed point sets.
The fixed point sets occur in two types - zero dimensional (isolated fixed points), which they called nuts after a canonical
example of a 'nut'-solution \cite{Newman:1963},  and two-dimensional, named bolts. Consequently, the gravitational instantons are classified according to
the number and type of nuts and bolts they possess. There exists as well a close relation between the fixed point sets structure and the topological invariants
of the solution, like Euler number and Hirzebruch signature,  which further supports the relevance of the classification.

\paragraph{} Chen and Teo \cite{Chen:2010} recently discussed  a refinement of this classification based on the  mathematical techniques developed
in \cite{Hollands:2007}, \cite{Hollands:2008}. In fact vacuum
gravitational instantons with $U(1)\times U(1)$ symmetry group can
be classified by simple extension of the classification theorems of
\cite{Hollands:2007}, \cite{Hollands:2008}\footnote{In the asymptotically flat  and Kaluza-Klein case
considered respectively in \cite{Hollands:2007} and
\cite{Hollands:2008}, there are canonical generators of the symmetry
group $U(1)\times U(1)$ coinciding with the standard Killing fields
and determined by the simple asymptotic geometry. In both cases the
corresponding classification theorems for the AE and AF
gravitational instantons are obtained directly from the
classifaction theorems \cite{Hollands:2007} and \cite{Hollands:2008}
with no horizon \cite{Chen:2010}. In the case of ALF instantons
where the asymptotic geometry is twisted there are no obvious
canonical generators for the symmetry group $U(1)\times U(1)$. In
ALF case  the gravitational instantons could be classified in the
spirit of  \cite{Hollands:2007} and \cite{Hollands:2008} in terms of
the interval structure and the asymptotic geometry, as conjectured
in \cite{Chen:2010}. }. These works proposed to consider the fixed
point set of the whole $U(1)\times U(1)$ isometry group in order to
describe axisymmetric solutions to the Einstein equation. In the
special case when the orbit space of the $U(1)\times U(1)$ isometry
group is parameterized by Weyl canonical coordinates $\rho$ and $z$,
the fixed point set represents a set of intervals along the z-axis,
hence it is called an interval structure. By definition, exactly one
Killing field vanishes along each of the intervals, so they
correspond to two-dimensional fixed point sets, or bolts. The
turning points, where two adjacent intervals intersect, represent
fixed points for the whole $U(1)\times U(1)$ isometry group. Thus,
they constitute a nut for any Killing field that generates isometries in the $U(1)\times U(1)$ group, which is linearly
independent with the directions of the two intersecting intervals.
In this way, the interval structure encodes the whole information
about the number and the position of the nuts and bolts.

\paragraph{} Another important classification of gravitational instantons is made on the basis of the asymptotic structure of spacetime \cite{Gibbons:1979gd}.
They divide into two major groups -  asymptotically locally Euclidean (ALE) and asymptotically locally flat (ALF). The asymptotically locally Euclidean instantons
are diffeomorphic  to $R \times (S^3/\Gamma)$ near infinity, where $\Gamma$ is a nontrivial discrete subgroup of $SO(4)$ with a free action on $S^3$, and the metric
asymptotically approaches the standard flat metric. The trivial case  when $\Gamma$  is the identity is called asymptotically Euclidean (AE). The asymptotically
locally flat instantons are diffeomorphic to the same manifold at infinity, however their metric tends to $ds^2 = dr^2 + r^2(\sigma^2_1 + \sigma^2_2) + \sigma^2_3$
instead, where $\{\sigma_1, \sigma_2, \sigma_3\}$ are the left-invariant one-forms on $S^3$. In the most common cases $\Gamma$ is the cyclic group $Z_n$.
Then, the spacial boundary at infinity of the ALF instantons represents a $S^1$ fibration over $S^2$ labeled by its first Chern class $c_1 = n$.
The case when $\Gamma$ is the identity, or, equvallently, $c_1 = 1$, is known as the Hopf fibration of $S^3$. The trivial case when $c_1=0$
is said to be asymptotically flat (AF), since it is topologically $S^1\times S^2$ at spacial infinity.

\paragraph{}The topological structure of infinity can be determined by the interval structure \cite{Hollands:2007}.
Suppose that the two semi-infinite intervals have directions $l_1 = (p_1, q_1)$ and $l_N = (p_N, q_N)$ with respect to a basic in $U(1)\times U(1)$, then the spacial boundary at infinity
is diffeomorphic to a lens space $L(q_1p_2-p_1q_2,w_1q_1-w_2p_1)$, where the integers $w_1$, $w_2$ satisfy the equation $w_1q_2 - w_2p_2 = ±1$.
Considering the properties of the lens spaces,  the following diffeomorphisms  $L(0, 1)\simeq S^1 \times S^2$, $L(1, 0)\simeq S^3$, and $L(p, 1)\simeq S^3/Z_p$
apply.

\section{5D Black Holes on ALF Gravitational Instanton}

\paragraph{}The 5D vacuum black holes on gravitational instanton can be described as solutions to the 5D Einstein equations which,
if the horizon is removed, will reduce to the direct product of a flat time dimension and the corresponding 4-dimensional Euclidean instanton.
In this way the instanton represents the spacial background of the solution and determines the topology of the spacetime at infinity.
In the current paper we will consider only ALF instantons with $U(1)\times U(1)$ group which possess a single bolt (or nut).
For simplicity we will also assume that the black hole is static. This restriction reduces the known black hole solutions to a couple of cases:
black hole on Taub-Nut and Taub-Bolt instantons. While the first is relatively well studied \cite{Cai:2006}, \cite{Ishihara:2008}, \cite{Ishihara:2007},
the second one was discovered only recently and we provide its first systematical investigation.

In our further investigations we will use extensively the interval structure of the solutions we consider. For that reason we will
describe it briefly. As already mentioned, the solution representing static black holes on ALF gravitational instanton possesses
a $R\times U(1)\times U(1)$ isometry groups generated by three commuting Killing fields. We will represent them in adapted coordinates in
the following way: $\xi = \frac{\partial}{\partial t}$ is the Killing field associated with time translations, $k  = \frac{\partial}{\partial\psi}$
is the Killing field corresponding to $S^1$ fibre, and $\zeta = \frac{\partial}{\partial\phi}$ is the remaining spacelike Killing vector.

\bigskip

\begin{figure} [h]
\begin{center}
      \includegraphics[width=12.cm]{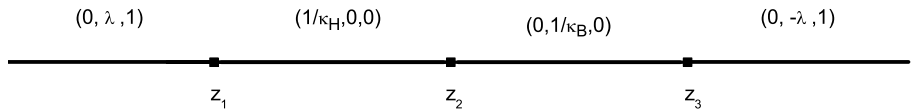}
\caption{Interval structure of black hole on a ALF gravitational instanton} \label{rodstr_inst}
        \end{center}
\end{figure}

Then, provided that the ALF instanton possesses a single bolt(or nut), the solution can be described by the interval structure
presented on fig. $\ref{rodstr_inst}$. The direction of each interval is normalized to the surface gravity associated to it.
Thus, $\kappa_H$ is the surface gravity of the horizon, and $\kappa_B$ is the surface gravity corresponding to a regular bolt.
The finite interval along the timelike Killing field corresponds to the horizon, and in the cases we consider it has spherical topology according to the topology theorem of \cite{Hollands:2007}.
The finite spacelike interval denotes the fixed point set of the isometry generated by the  Killing field corresponding to the $S^1$ fibre at infinity.
According to the instanton classification terminology, this is a bolt. The solution describing black hole on Taub-Nut instanton is recovered
as the particular limit when this interval collapses to a point. The two semi-infinite intervals are directed along the linear combinations of
the spacelike Killing fields $l_L = \eta + \lambda k$ and $l_R = \eta - \lambda k$,  which are generators of the $U(1)\times U(1)$ group. The parameter $\lambda$ is connected with the length of the $S^1$ fibre at infinity $L$ as $\lambda = L/4\pi$. According to the rule for determining the topology of the spacelike sections at infinity which we mentioned in
the previous section, such semi-infinite interval alignment corresponds to a $L(1,0)$ topology, of equivalently a Hopf fibration of $S^3$. In the case of
black hole on Euclidean Scwarzchild instanton the parameter $\lambda$ vanishes leading to a trivial $S^1\times S^2$ topology of the spacelike
boundary at infinity.

\section{Thermodynamic Properties of Black Holes on ALF Gravitational Instantons}

\subsection{Nut charge and nut potential}

\paragraph{}A distinctive feature of the ALF solutions is that they possess an additional charge,
the so called nut charge, which is absent in the asymptotically flat
case. The nut charge is determined by the 'twist' of the $S^1$
bundle over the two-sphere at infinity. It is proportional to its
first Chern class and therefore vanishes if spacial infinity is
diffeomorphic to the trivial bundle $S^1\times S^2$.  The nut charge
can be defined by a Komar-like integral \cite{Hunter:1998}

\begin{equation}
N = - {1\over 8\pi} \int_{C^2} d\left(\frac{k}{V}\right),
\end{equation}
where $k$ is the Killing 1-form associated with the $S^1$ fibre at infinity,  $V$ is its norm and $C^2$ is a two-dimensional surface, encompassing the nut. If the solution possesses a non-trivial geometry this quantity does not vanish, as one may otherwise expect by Stokes' theorem. The reason is that the 1-form $k/V$ is not globally defined. Two coordinate charts should be introduced in order to obtain a complete atlas for the manifold, which lead to a non-trivial value of the integral. This feature was originally observed by Misner investigating the Taub-Nut solution \cite{Misner:1963} and the coordinate singularity which typically arises in similar solutions was called 'Misner string'.

If we consider the interval structure presented on fig. $\ref{rodstr_inst}$ with semi-infinite intervals directed along
the Killing vectors $l_L = \eta + \lambda k$ and $l_R = \eta - \lambda k$, we can express the nut charge in the following way

\begin{eqnarray}
N = - {1\over 4} \int_{C^{2}/U(1)}i_\eta d\left(\frac{k}{V}\right)=
-{1\over 4} \int_{C^{2}/U(1)}\left({\cal L}_{\eta} - d i_\eta\right)\frac{k}{V}=
 {1\over 4}\left[\left(\frac{i_\eta k}{V}\right)_{I_R} - \left(\frac{i_\eta k}{V}\right)_{I_L}\right],
\end{eqnarray}
where we have denoted the semi-infinite intervals left and right to the nut with $I_L$ and $I_R$.
We recognize that the Killing vectors $l_L = \eta + \lambda k$ and $l_R = \eta - \lambda k$ vanish on the left and right semi-infinite intervals
respectively and obtain as a final result

\begin{equation}\label{Nut}
N = {1\over 2}\lambda = \frac{L}{8\pi}.
\end{equation}
This simple relation can be demonstrated in the case of Taub-NUT or Taub-Bolt solutions where the length of the $S^1$ fibre at infinity is $L = 8\pi n$.
Then it yields the well known value of the nut charge $N = n$.

\paragraph{}In addition to the nut charge there exists another related characterstic, called a nut potential.
This is revealed if we examine the 1-form $i_\xi i_k \star dk$

\begin{eqnarray}
d i_\xi i_k \star d k &=& \left({\cal L}_\xi - i_\xi d\right)i_k \star d k = - i_\xi d i_k \star d k = \nonumber \\
&=&- i_\xi\left({\cal L}_k - i_k d\right)\star d k = i_\xi i_k d \star d k = 2 i_\xi i_k \star R(k) = \nonumber \\
&=&2\star\left[R(k)\wedge k \wedge \xi \right]
\end{eqnarray}

In theories satisfying $R(k)\wedge k \wedge \xi = 0$, which includes
the vacuum case and the static  sector of Einstein-Maxwell-dilaton
gravity, the 2-form $d i_\xi i_k \star d k$ vanishes. The 1-form
$i_\xi i_k \star d k$ is invariant under the Killing fields $ \xi,
k$ and $\eta$ and can be viewed as defined on the factor space $
\hat{M}= M/ R \times U(1)^2$.  Since the factor space $ \hat{M}= M/ R
\times U(1)^2$ is simply connected \cite{Hollands:2007}, there exists a globally defined
potential $\chi$, such as

\begin{equation}\label{Nut potential}
i_\xi i_k \star d k = d\chi.
\end{equation}

This relation can be regarded as a definition of the nut potential. The existence of the nut potential has been noticed also previously \cite{Gibbons:1979c}
by different means considering 4D Euclidean instanton solutions and performing dimensional reduction of the field equations along the Killing field $k$.
From the definition we can deduce some useful properties of the nut potential. It is constant on the fixed point set of the Killing field $k$,
and on the black hole horizon as well, provided that the horizon is bifurcational.

\subsection{Mass, Tension and Smarr-like Relations}

\paragraph{}The definition of conserved charges for non-asymptotically flat spacetimes is still an open area of research.
The classical approach by Brown and York \cite{Brown:1993}, which is well defined in the asymptotically flat case, is known
to encounter some difficulties in the general case. In the most cases non-flat asymptotics lead to divergent boundary stress-energy tensor.
There exists a regularization procedure considering the boundary as embedded in some appropriate reference spacetime with the same asymptotics and calculating the stress-energy tensor with respect to this reference spacetime. The choice of reference spacetime is however ambiguous, moreover it is not always possible to embed a boundary with arbitrary induced metric into the reference background.

\paragraph{} Another procedure for obtaining a regular gravitational action, which is developed in the literature, is adding a term, called a counterterm,
to the boundary at infinity \cite{Kraus: 1999}. By definition, it is a functional only of curvature invariants of the induced metric on the boundary,
therefore has no influence on the equations of motion, and should be chosen appropriately in order to cancel the divergences.
Mann and Stelea \cite{Mann:2005} proposed the  following counterterm (see also Astefanesei and Radu \cite{Radu:2006})
\begin{equation}\label{counterterm}
I_{ct} = {1\over 8\pi} \int d^4x \sqrt{-h} \sqrt{2{\cal R}},
\end{equation}
which is relevant for the asymptotic we  consider. It leads to the boundary stress-energy tensor
\begin{equation}\label{SET}
T_{ij} = {1\over 8\pi} \left[ K_{ij} - K h_{ij} - \Psi ({\cal R}_{ij} - {\cal R}h_{ij} ) - h_{ij} D^kD_{k}\Psi + D_{i}D_{j}\Psi   \right],
\end{equation}
where  K is the trace of the extrinsic curvature $K_{ij}$ of the boundary, R and $D_k$ are the Ricci scalar and the covariant derivative
with respect to the boundary metric $h_{ij}$, and $\Psi= \sqrt{2\over {\cal R}}$. If the boundary geometry has an isometry generated by the Killing vector $\xi$,
the conserved charge associated to it is given by

\begin{equation}
{\cal Q} = \int_{\Sigma} d\Sigma_{i} T^{i}_{j}\xi^{j},
\end{equation}
which represents the mass in the case when $\xi = \partial/\partial t $, and the tension, when $\xi = \partial/\partial \psi$ \cite{Stelea:2009}.

In our investigation we will consider an independent definition of
the conserved charges by the Komar-like  integrals
\cite{Townsend:2001}, or more explicitly, their generalized form
introduced in  \cite{YN1}. In spacetimes possessing a spacelike
Killing vector corresponding to a compact dimension at infnity the
ADM mass and the tension can be expressed by the generalized Komar
integrals

\begin{eqnarray}\label{MT}
M_{ADM} &=&  - {L\over 16\pi} \int_{S^{2}_{\infty}} \left[2i_k \star d\xi - i_\xi \star d k \right] \\
{\cal{T}} &=&  - {1\over 16\pi} \int_{S^{2}_{\infty}} \left[i_k \star d\xi - 2i_\xi \star d k \right], \nonumber
\end{eqnarray}

where $\xi = \frac{\partial}{\partial t}$ is the Killing field
associated with time translations, $k  =
\frac{\partial}{\partial\psi}$ is the Killing field corresponding to
the compact dimension, $L$ is the length of the $S^1$ fibre  and
$S^2_{\infty}$ is the base space of $S^1$-fibration at infinity. Let
us consider the expression for the ADM mass. We can reduce it to the
factor space $\hat{M}$ by acting with the Killing field $\eta$
associated with the azimuthal symmetry of the two-dimensional sphere
at infinity.

\begin{eqnarray}\label{M_ADM}
M_{ADM} =   {L\over 8} \int_{Arc(\infty)} \left[2i_\eta i_k \star d\xi - i_\eta i_\xi \star d k \right],
\end{eqnarray}
The integration is now performed over the semicircle representing the boundary of the two-dimensional factor space at infinity.
Using the Stokes' theorem the integral can be further expanded into a bulk term over $\hat{M}$ and an integral over the rest of the boundary
of the factor space which is represented by the interval structure $I_i$.

\begin{eqnarray}
M_{ADM} =   {L\over 8} \int_{\hat{M}} \left[2di_\eta i_k \star d\xi
- d i_\eta i_\xi \star d k \right] - {L\over 8}\sum_i \int_{I_i}
\left[2i_\eta i_k \star d\xi - i_\eta i_\xi \star d k \right] ,
\end{eqnarray}

First we are going to consider the bulk integral

\begin{eqnarray}
 {L\over 8} \int_{\hat{M}} \left[2di_\eta i_k \star d\xi - d i_\eta i_\xi \star d k \right] &=&  {L\over 8} \int_{\hat{M}} \left[2i_\eta i_k d\star d\xi - i_\eta i_\xi d\star d k \right] = \nonumber \\
&=&  {L\over 4} \int_{\hat{M}} \left[2i_\eta i_k \star R(\xi) -
i_\eta i_\xi \star R(k) \right],
\end{eqnarray}
where we have used the Ricci-identity $d\star d K=2\star R(K)$ applying for any Killing field $K$.
It follows that in the vacuum case the bulk term vanishes and the ADM mass can be expressed only by means of the contributions on the interval structure.
If we consider the interval structure presented on fig. $\ref{rodstr_inst}$ we obtain

\begin{eqnarray}\label{Smarr}
M_{ADM} =  &-& {L\over 8} \int_{I_H} \left[2i_\eta i_k \star d\xi - i_\eta i_\xi \star d k \right] - {L\over 8} \int_{I_B} \left[2i_\eta i_k \star d\xi - i_\eta i_\xi \star d k \right]- \nonumber \\
 &-& {L\over 8} \int_{I_L \bigcup I_R} \left[2i_\eta i_k \star d\xi - i_\eta i_\xi \star d k \right].
\end{eqnarray}

The integral over the horizon interval $I_H$ is interpreted as the
intrinsic mass of the black hole. We will consider only
bifurcational horizons, meaning the Killing field $\xi$ vanishes on them, which leads to a simpler expression for the black hole
mass

\begin{eqnarray}\label{MH}
M_{H} =  - {L\over 4} \int_{I_H} i_\eta i_k \star d\xi
\end{eqnarray}

In the similar way the integral over the bolt $I_B$ is assumed to express the intrinsic mass of the instanton.
By definition the Killing field $k$ vanishes on the bolt leading to

\begin{eqnarray}\label{MB}
M_{B} =   {L\over 8} \int_{B} i_\eta i_\xi \star d k
\end{eqnarray}

We can define a surface gravity of the bolt $\kappa_B$ as it was introduced in the AF case for the Euclidean Schwarzchild instanton \cite{Kastor:2008}
by the relation
\begin{eqnarray}
d k = 2\kappa_B n_1\wedge n_2
\end{eqnarray}
where $n_1$ and $n_2$ are the spacelike unit normals to the bolt, in adition to the timelike one $n_\xi = \frac{\xi}{\sqrt{|g( \xi, \xi)|}}$.
Using the bolt surface gravity and the generalized bolt area defined as

\begin{eqnarray}
\tilde{A}_B = \int_{Bolt} \sqrt{|g(\xi, \xi)|}d A,
\end{eqnarray}
where $d A$ is the bolt surface element, we can provide another equivalent expression for the instanton mass

\begin{eqnarray}
M_{B} &=&  {L\over 16\pi} \int_{B} i_\xi \star d k = {L\over 16\pi} \int_{B} \star \left(d k \wedge \xi\right) = \\
&=& {L\over 16\pi} \int_{B} i_{n_1} i_{n_2} i_{n_\xi} \left( dk\wedge \xi\right)d A = \frac{L}{8\pi}\kappa_B \tilde{A}_B.
\end{eqnarray}

Let us go back to equation ($\ref{Smarr}$) and investigate the meaning of the remaining integrals along the semi-infinite intervals $I_L$ and $I_R$.
According to the interval structure on fig. $\ref{rodstr_inst}$ the linear combinations $l_L = \eta +  \lambda k$ and $l_R = \eta -  \lambda k$ vanish on $I_L$
and $I_R$ respectively, for that reason it is satisfied that $i_\eta i_k \star d\xi\mid_{I_{L/R}} = 0$. Consequently, we have

\begin{eqnarray}
-{L\over 8} \int_{I_L} \left[2i_\eta i_k \star d\xi - i_\eta i_\xi \star d k \right] &=& {L\over 8} \int_{I_L}i_\eta i_\xi \star d k  = \\
 &=&{L\over 8}\lambda\int_{I_L}i_\xi i_k \star d k = {L\over 8} \lambda \int_{I_L} d\chi,
\end{eqnarray}
and the same expression with  a minus sign is valid for the right semi-infinite interval.

In this way we obtain the relation

\begin{eqnarray}
M_{ADM} &=& M_H + M_{B} + {L\over 8} \lambda \int_{I_L} d\chi - {L\over 8} \lambda \int_{I_R} d\chi = \\
 &=& M_H + M_{B} + {L\over 8}\lambda\left( \chi(z_1) - \chi(\infty)\right) - {L\over 8}\lambda\left(\chi(\infty) - \chi(z_3)\right)
\end{eqnarray}
We have denoted by $z_1$ and $z_3$ the end points of the left and right semi-infinite intervals, in which the first one intersects with the
horizon and the other with the bolt (see fig.$\ref{rodstr_inst}$). Since the nut potential is constant  both  on the horizon and the bolt,
we have for continuity reasons that $\chi(z_1) = \chi(z_3) \equiv \chi$. Considering the definition of the nut charge ($\ref{Nut}$)
and normalizing the nut potential so that it vanishes at infinity, we obtain a Smarr-like relation in the form

\begin{eqnarray}
M_{ADM} = M_H + M_{B} + {L\over 4}\lambda \chi= M_H + M_{B} + {L\over 2} N\chi.
\end{eqnarray}

Performing similar operations on the tension expression ($\ref{MT}$) we can derive the following Smarr-like law for the tension

\begin{equation}\label{ST}
{\cal{T}}L = \frac{1}{2}M_H + 2M_{B} + L N\chi.
\end{equation}

\subsection{Euclidean action and entropy}

The Euclidean gravitational action for a metric $g$ and matter fields $\phi$ is defined as
\begin{equation}
I(g,\phi) = -\int_M \left[ {R\over 16\pi}  +L_m(g, \phi)\right]-{1\over 8\pi}\oint_{\partial M} K
\end{equation}
where $R$ is the scalar curvature of $g$ and $L_m$ is the matter Lagrangian. The integration is performed over a section $M$ of the complexified spacetime,
where the metric is real and positive-definite and $K$ is the trace of the extrinsic curvature of the boundary $\partial M$. The action is well defined for
spatially compact geometries, but diverges for noncompact ones. As already mentioned, to define a finite action for noncompact geometries, one should either
choose a reference background and consider the action with respect to it, or add a suitable counterterm. In our approach we use the counterterm ($\ref{counterterm}$) which is consistent with the asymptotical structure of the solutions we investigate. In this way the gravitational action transforms to

\begin{equation}\label{action}
I(g,\phi) = -\int_M \left[ {R\over 16\pi}  + L_m(g, \phi)\right]-{1\over 8\pi}\oint_{\partial M} \left[K - \sqrt{2{\cal R}}\right].
\end{equation}

The Euclidean action is connected to the partition function $Z$ of a gravitational system \cite{Gibbons:1976} as

\begin{equation}
Z = \int D(g)D(\phi)e^{iI(g, \phi)},
\end{equation}
where $D(g)$ is a measure on the space of metrics $g$, and $D(\phi)$ is a measure on the space of matter fields $\phi$.
The main contribution to the path integral will come from metric and matter fields $g_0$ and $\phi_0$ which are solutions to the field equations.
Thus, in a leading order approximation we obtain

\begin{equation}
\ln{Z} = i I(g_0, \phi_0).
\end{equation}

By thermodynamical arguments, the partition function for a grand canonical ensemble at temperature $T$ is connected to its thermodynamical
potential ${\cal{F}}= M - TS- \sum{\mu_iC_i}$ by

\begin{equation}
\ln{Z} = - {\cal{F}}T^{-1},
\end{equation}
where $\mu_i$ are the chemical potentials associated with the conserved quantities $C_i$. This relation allows to determine the entropy $S$ of a
gravitational system by means of its gravitational action and will be used in the investigation of particular solutions in sections 5 and 6.

\section{Thermodynamics of particular solutions}

\subsection{Black hole on Taub-Bolt Instanton}

The solution describing a static black hole on Taub-Bolt instanton was recently discovered by Chen and Teo \cite{Chen:2011}.
It has the following metric in C-metric-like coordinates\footnote{The relationship of the used C-metric-like coordinates to Weyl canonical coordinates is discussed in Appendix H of \cite{Harmark:2004}.}

\begin{eqnarray}
{ds}^{2}&=& - {\frac {  1+cy }{1+cx}}\,{dt}^{2}+ {\frac {F ( x,y ) }{H ( x,y ) }} \left( {d\psi}+\Omega d\phi \right) ^{2} \nonumber \\
&&+\frac{2{\kappa}^{4} \left( 1-c \right)  \left( 1+cx \right) H ( x,y ) }{ \left( 1-{\alpha}^{2} \right) \left( x-y \right) ^3} \left( {\frac {{dx}^{2}}{G ( x ) }}-{\frac {{dy}^{2}}{G ( y ) }}+A\,{{d\phi}}^{2} \right)
\end{eqnarray}
where $\Omega$ and $A$ are defined as
\begin{eqnarray}
\Omega&=&{\frac {2 \alpha\,{\kappa}^{2} [2+x+y+c \left( 1+x
 \right)  \left( 1+y \right) ] }{ \left( 1-{\alpha}^{2}
 \right)  \left( x-y \right) }} , \nonumber \\
A&=&-{\frac {2 \left( 1+x \right)  \left( 1+y \right) }{ \left( 1-c
 \right)  \left( x-y \right) }}\,
\end{eqnarray}
and the functions $G(x)$, $H(x,y)$ and $F(x,y)$ have the form
\begin{eqnarray}
G ( x ) &=& \left( 1+cx \right)  ( 1-{x}^{2} )\,,\cr
H ( x,y ) &=& \left( 1+cx \right)  \left( 1-y \right)^{2}-{
\alpha}^{2} \left( 1+cy \right)  \left( 1-x \right)^{2},  \\
F ( x,y ) &=& ( 1-{\alpha}^{2} )  \left( 1-x
 \right)  \left( 1-y \right)  \left( 1+cx \right). \nonumber
\end{eqnarray}

The parameters $\kappa$, $c$, $\alpha$ and coordinates $t$, $x$, $y$ take the ranges $\kappa>0$, $0\leq c < 1$, $\alpha^2<1$, $-\infty<t<\infty$, $-1\leq x \leq 1$, $-\frac{1}{c}\leq y\leq -1$. Physical infinity corresponds to $(x,y)=(-1,-1)$.

\paragraph{}The interval structure of solution consists of
\begin{itemize}
\item a semi-infinite space-like interval located at $(x=-1,-\frac{1}{c}\leq  y< -1)$, with direction $l_L=(0,{\frac {2 \alpha{\kappa}^{2}}{1-{\alpha}^{2}}},1)$;
\item a finite timelike interval  located at $(-1\leq x\leq 1,y=-\frac{1}{c})$ with direction $l_H=\frac{1}{\kappa_H}(1,0,0)$, which corresponds to a horizon;
\item a finite space-like interval located at $(x=1,-\frac{1}{c}\leq y\leq -1)$ with direction $\l_B=\frac{1}{\kappa_B}(0,1,0)$, which corresponds to a bolt;
\item a semi-infinite space-like interval located at $(-1<x\leq 1,y=-1)$ with direction $l_R=(0,-{\frac {2 \alpha{\kappa}^{2}}{1-{\alpha}^{2}}},1)$.
\end{itemize}

The directions of the finite intervals are normalized to the surface gravity corresponding to them which is equal to

\begin{equation}
\kappa = \sqrt{\mp{1\over 2}\nu_{\alpha;\beta}\nu^{\alpha;\beta}}\mid_{interval}
\end{equation}
for a timelike and spacelike interval respectively, directed along a Killing vector $\nu$. The surface gravity on the black hole horizon is computed as
\begin{equation}\label{temp}
\kappa_H =\frac{1}{2\kappa^2}\sqrt{\frac{1-\alpha^2}{2c(1+c)}},
\end{equation}
and the surface gravity on the bolt takes the value

\begin{equation}
\kappa_B =\frac {1-{\alpha}^{2}}{2{\kappa}^{2}\sqrt {1-{c}^{2}}} = \frac{2\pi}{L},
\end{equation}
where $L$ is the length of the $S^1$ fibre at infinity. The last equality is imposed in order to avoid conical singularities. To obtain a regular solution we should impose a further condition $\alpha = \frac{\sqrt{1-c^2}}{2}$, which ensures that the solution is free of orbitfold singularities \cite{Hollands:2007}, \cite{Chen:2011}.

In the limit $c = 0$ the timelike interval vanishes and we recover the background spacetime which represents a Taub-Bolt instanton
trivially embedded in 5-dimensional spacetime, also called in the literature Kaluza-Klein monopole \cite{Gross:1983}, \cite{Sorkin:1983}.
If we perform the following coordinate transformation and redefinition of the parameters

\begin{equation}\label{parameter}
r={\frac {{\kappa}^{2} [1-y-{\alpha}^{2} \left( 1-x \right) ] }{ \left( 1-{\alpha}^{2} \right)  \left( x-y \right) }}\,,\quad
\cos\theta=\frac {2+x+y}{x-y} \,,\quad
m=\frac{\kappa^2}{2}\frac{1+\alpha^2}{1-\alpha^2}\,,\quad
n={\frac {{\alpha \kappa}^{2}}{1-{\alpha}^{2}}}\,,
\end{equation}
the standard form of its metric is obtained

\begin{eqnarray}
ds^2 &=& - dt^2 +  \frac{\left(r^2 + n^2- 2mr\right)}{\left(r^2 - n^2\right)}\left(d\psi + 2n\cos{\theta}d\phi\right)^2+ \nonumber \\
&+& \frac{\left(r^2 - n^2\right)}{\left(r^2 + n^2 - 2mr \right)}dr^2 + \left(r^2 - n^2 \right) \left(d\theta^2 + \sin^2{\theta}d\phi^2\right)
\end{eqnarray}

\paragraph{} The solution describing black hole on a Taub-Bolt instanton possesses the following asymptotical behavior at $(x, y)\rightarrow (-1,-1)$

\begin{eqnarray}
ds^2 &=& - \left(1 - \frac{2\kappa^2 c}{r}\right)dt^2 + \left( 1 - \frac{2m}{r} + \frac{\kappa^2 c}{r}\right)\left(d\psi + 2n\cos{\theta}d\phi\right)^2+ \nonumber \\
&+&\left(1 + \frac{2\kappa^2c}{r}\right)\left[\frac{dr^2}{\left( 1 - \frac{2m}{r} + \frac{\kappa^2 c}{r}\right)} + r^2 \left(d\theta^2 + \sin^2{\theta}d\phi^2\right)\right],
\end{eqnarray}
where $r$ and $\theta$ are asymptotical coordinated related to the C-metric coordinates as

\begin{eqnarray}\label{asympt}
x = -1 + 2\kappa^2\left(1-c\right)\cos^2{\left({\theta\over 2}\right)} \frac{1}{r} + O(\frac{1}{r^2}), \nonumber\\
y = -1 - 2\kappa^2\left(1-c\right)\sin^2{\left({\theta\over 2}\right)} \frac{1}{r} + O(\frac{1}{r^2})
\end{eqnarray}

Using the asymptotical metric we can compute the ADM mass and the tension by evaluating the Komar integrals ($\ref{MT}$). Thus we obtain

\begin{eqnarray}\label{MA}
M_{ADM} &=& \frac{L}{2}\left( m + \frac{3}{2}c \kappa^2\right), \nonumber \\
{\cal{T}} &=& m,
\end{eqnarray}
where the parameter $m$ is defined by ($\ref{parameter}$).
The mass and the tension of the solution can be calculated also in a different way, applying the counterterm  method we described in section 4.2.
Direct calculation of the relevant components of the boundary stress-energy tensor ($\ref{SET}$) leads to the result

\begin{eqnarray}\label{SET1}
8\pi T_t^t &=& \frac{1}{r^2}\left(m + \frac{3}{2}c \kappa^2\right) + O(\frac{1}{r^3}),  \\
8\pi T_\psi^\psi &=& \frac{2m}{r^2} + O(\frac{1}{r^3}), \nonumber
\end{eqnarray}
hence

\begin{eqnarray}
M_{ADM} &=& \frac{1}{8\pi}\int\left(m + \frac{3}{2}c \kappa^2\right)\sin{\theta}d\theta d\phi d\psi, \nonumber \\
{\cal{T}} &=& \frac{1}{4\pi}\int m\sin{\theta}d\theta d\phi.
\end{eqnarray}

It is obvious that performing the integration yields the same expression as previously obtained by the Komar approach. In the limit $c=0$
we recover the mass of the Kaluza-Klein monopole, which was calculated in \cite{Mann:2005}.

The intrinsic masses of the black hole and the instanton are obtained directly using ($\ref{MH}$) and ($\ref{MB}$)

\begin{eqnarray}
M_H &=& L c\kappa^2, \nonumber \\
M_B &=& {L\over 4}\kappa^2\left(1-c\right).
\end{eqnarray}

Further, we compute the one-form ($\ref{Nut potential}$) defining the nut potential along the left semi-infinite interval $(x=-1,-\frac{1}{c}\leq  y< -1)$

\begin{equation}
d\chi_L = \frac{4\alpha\left(1-\alpha^2\right)(1-y)\left[ \left(1+c\right) + \left(1+cy\right)\right]}{(1-c)\left[(1-y)^2 - \left(1+c\right)\left(1+cy\right)\right]^2}dy,
\end{equation}
which yields after integration along the interval

\begin{equation}
\chi_L = \frac{4\alpha\left(1-\alpha^2\right)\left(1+cy\right)}{(1-c)\left[(1-y)^2 - \left(1+c\right)\left(1+cy\right)\right]} + \chi_0.
\end{equation}

We determine the integration constant $\chi_0$ from the requirement that the nut potential vanishes at infinity obtaining

\begin{equation}
\chi_0 = \frac{4\alpha(1-\alpha^2)}{(3+c^2)} = \alpha,
\end{equation}
where we have used the regularization condition $\alpha = \frac{\sqrt{1-c^2}}{2}$.

At the turning point $y = -{1\over c}$ where the left semi-infinite interval intersects with the horizon interval the nut potential takes a
simple value $\chi_L(-{1\over c}) = \alpha$.

In a similar way we calculate the nut potential along the right semi-infinite interval $(-1<x\leq 1,y=-1)$ obtaining

\begin{eqnarray}
d\chi_R &=& \frac{4\alpha\left(1-\alpha^2\right)(1-c)(1-x)\left[ \left(1+c\right) + \left(1+cx\right)\right]}{\left[4(1+cx) - \alpha^2(1-c)\left(1-x\right)^2\right]^2}dx, \\
\chi_R &=&\frac{\alpha\left[ (1-c)\left(1-x\right)^2 - 4(1+cx)\right]}{\left[\alpha^2(1-c)\left(1-x\right)^2 - 4(1+cx)\right]}.
\end{eqnarray}

At the turning point $x = 1$ where the right semi-infinite interval intersects with the bolt the nut potential takes the value $\chi_R(1) = \alpha$.
As expected, the values of the nut potential at the turning points where the semi-infinite intervals intersect the finite intervals coincide, since it is constant on the black hole horizon and the bolt.

The nut charge is obtained immediately from ($\ref{Nut}$) as

\begin{equation}
N = \frac {\alpha{\kappa}^{2}}{1-{\alpha}^{2}}=n
\end{equation}

Thus we obtain the Smarr-like relation for the mass

\begin{equation}
M_{H} + M_{B}+ \frac{L}{2}N\chi = \frac{L}{2}\left( 2 c\kappa^2 + \frac{\kappa^2}{2}(1-c) + \frac {\alpha^2{\kappa}^{2}}{1-{\alpha}^{2}}\right) = \frac{L}{2}\left(m+ \frac{3}{2}\kappa^2c\right)
\end{equation}
which is obviously consistent with the expression of the $M_{ADM}$ mass we obtained by direct calculation of the Komer integral or by the counterterm method.
It is easy to verify that the Smarr-like relation for the tension ($\ref{ST}$) is also fulfilled.

\paragraph{}In the following investigation we will compute the Euclidean action corresponding to the solution as it was described in section 4.3.
Considering equation  ($\ref{action}$) and assuming that the field equations in vacuum are satisfied it follows that only the surface term in the
gravitational action will contribute. We obtain for the extrinsic curvature and the Ricci scalar on the boundary

\begin{eqnarray}
K &=& \frac{2}{r}\left(1 -\frac{1}{2}\frac{m}{r}-\frac{5}{4}\frac{\kappa^2 c}{r}\right), \nonumber \\
\cal{R} &=& \frac{2}{r^2}\left(1 - 2\frac{\kappa^2 c}{r}\right),
\end{eqnarray}
which leads to the Euclidean action

\begin{equation}
I = i\beta\frac{L}{2}\left(m + \frac{1}{2}\kappa^2c\right).
\end{equation}
We have denoted with $\beta$ the period of the imaginary time coordinate which equals the inverse of the black hole temperature. Thus we obtain
for the thermodynamical potential
\begin{equation}
{\cal{F}} = M - TS = \frac{L}{2}\left(m + \frac{1}{2}\kappa^2c\right).
\end{equation}
Considering the expression for the ADM mass ($\ref{MA}$) the last expression yields

\begin{equation}\label{TS}
TS = \frac{L}{2}\kappa^2c.
\end{equation}
The temperature of the black hole is computed by the surface gravity on the horizon according to

\begin{equation}\label{temp}
T =\frac{\kappa_H}{2\pi} = \frac{1}{4\pi\kappa^2}\sqrt{\frac{1-\alpha^2}{2c(1+c)}}.
\end{equation}

On the other hand, the black hole area can be determined by straightforward integration as

\begin{eqnarray}\label{area}
A_H = \int_H \sqrt{g_H}dx d\phi d\psi = 8\pi L \kappa^4 c \sqrt{\frac{2c(1+c)}{1-\alpha^2}}.
\end{eqnarray}

Comparing relations ($\ref{TS}$), ($\ref{temp}$) and ($\ref{area}$) we obtain that the entropy of the solution obeys

\begin{equation}
S = 2\pi L \kappa^4 c \sqrt{\frac{2c(1+c)}{1-\alpha^2}} = \frac{A_H}{4}.
\end{equation}

\subsection{Black hole on Taub-Nut Instanton}

The solution representing a charged black hole on a Taub-Nut instanton was found originally by Ishihara and Matsuno \cite{Ishihara:2005},
although it was called a Kaluza-Klein black hole with squashed horizon. Some of its thermodynamical properties were examined in \cite{Cai:2006}, \cite{Ishihara:2007}, \cite{Yazadjiev:2006} and the results partially overlap with our investigations. Note, however, that in the present discussion we choose to redefine the solution parameters such as the radius of the $S^1$ bundle at infinity to be normalized to unity.

The metric describing a neutral black hole on a Taub-Nut instanton is given by

\begin{eqnarray}\label{RHOM}
ds^2 &=& -\left(1 - \frac{\rho_{+}}{\rho}\right) d t^2 + {\rho + \rho_0 \over \rho - \rho_{+}} d\rho^2 + \rho \left( \rho + \rho_0 \right) \left(d\theta^2 + \sin^2\theta d\phi^2\right) + \nonumber \\
&&{\rho \over \rho + \rho_0 } \left(d\psi + r_{\infty} \cos\theta d\phi\right)^2,
\end{eqnarray}
where $r_{\infty} = \sqrt{\rho_0 \left( \rho_0 + \rho_{+} \right)}$.

\paragraph{} The interval structure associated to the solution consists of

\begin{itemize}
\item a semi-infinite space-like interval located at $\left( \rho \geq \rho_+, \theta = \pi \right)$ with direction $l_L = (0, r_{\infty}, 1)$;
\item a finite timelike interval located at $\left( \rho = \rho_+, 0 \leq\theta \leq\pi \right)$ with direction $l_H=\frac{1}{\kappa_H}(1,0,0)$ corresponding to the black hole horizon;
\item a semi-infinite space-like interval at $\left( \rho \geq \rho_+, \theta = 0 \right)$ with direction $l_R = (0, -r_{\infty}, 1)$.
\end{itemize}

The length of the $S^1$ fibre at infinity is equal to $L = 4\pi  \sqrt{\rho_0 \left( \rho_0 + \rho_{+} \right)} = 4\pi r_\infty$, and $\kappa_H$ is the surface gravity of the horizon.

The ADM mass and the tension of the solution are computed using the Komar integrals ($\ref{MT}$) as

\begin{eqnarray}\label{MN}
M_{ADM} &=&  \frac{L}{2}\left(\rho_{+} + {1\over 2}\rho_{0}\right)   \\
{\cal{T}} &=& \frac{1}{2} \left( \rho_{0} + {1\over 2}\rho_{+}\right). \nonumber
\end{eqnarray}

Again, we will calculate the asymptotic mass and the tension independently applying the counterterm method. The corresponding components of the
stress-energy tensor are found in the form

\begin{eqnarray}
8\pi T^{t}_{t} &=& {1\over \rho^2} \left({1\over 2}\rho_{0} + \rho_{+} \right) + {\cal O}({1\over \rho^3}),   \\
8\pi T^{\psi}_{\psi} &=& {1\over \rho^2} \left(\rho_{0} + {1\over 2}\rho_{+} \right) + {\cal O}({1\over \rho^3}). \nonumber
\end{eqnarray}

hence

\begin{eqnarray}
M_{ADM} &=& \frac{1}{8\pi}\int\left({1\over 2}\rho_{0} + \rho_{+} \right)\sin{\theta}d\theta d\phi d\psi ,  \\
{\cal{T}} &=& \frac{1}{8\pi}\int\left(\rho_{0} + {1\over 2}\rho_{+} \right)\sin{\theta}d\theta d\phi. \nonumber
\end{eqnarray}

After performing the integration we obtain the same expression for the ADM mass and the tension as by the Komar approach.

According to ($\ref{MH}$), the intrinsic mass of the black hole is found to be

\begin{eqnarray}
M_H = \frac{L}{2}\rho_+.
\end{eqnarray}

 Further, we compute the one-form ($\ref{Nut potential}$), which defines the nut potential along the left
 semi-infinite interval $\left( \rho \geq \rho_+, \theta = \pi \right)$
\begin{eqnarray}
 d\chi = -  \frac{L}{4\pi\left(\rho + \rho_0\right)^2} d\rho,
\end{eqnarray}
Note that it will have exactly the same form on the right semi-infinite interval. Integrating we obtain
\begin{eqnarray}
\chi &=& \frac{L}{4\pi\left(\rho + \rho_0\right)},
\end{eqnarray}
where we have taken into account that the integration constant should be chosen in such a way that the nut potential vanishes at infinity.
At the turning point $\rho = \rho_+$, where the left semi-infinite interval intersects with the horizon the nut potential takes the value $\chi(\rho_+) = \frac{4\pi}{L}\rho_0$.

Applying equation $\left(\ref{Nut}\right)$, we obtain the nut charge $N = \frac{L}{8\pi}= \frac{r_\infty}{2}$.
It is easy to show that the calculated quantities satisfy the Smarr-like relations for the mass and the tension we derived in section 4.2.

\paragraph{}The Euclidean action of the solution is calculated in the same way as demonstrated in the previous section for the black hole on Taub-Bolt solution obtaining

\begin{eqnarray}
I =  i\beta\frac{L}{4}\left( \rho_0 + \rho_+ \right) .
\end{eqnarray}

Consequently, the thermodynamical potential is

\begin{eqnarray}\label{FN}
{\cal{F}} = M - TS = \frac{L}{4}\left( \rho_0 + \rho_+ \right),
\end{eqnarray}

and the temperature and area of the black hole are determined by

\begin{eqnarray}\label{TN}
T &=&  \frac{1}{L}\sqrt{\frac{\rho_0}{\rho_+}} \\ \nonumber
A_H &=&  L^2\rho_+\sqrt{\frac{\rho_+}{\rho_0}}.
\end{eqnarray}

Considering expression ($\ref{MN}$), ($\ref{FN}$) and ($\ref{TN}$) we obtain the entropy of the solution

\begin{eqnarray}
S = \frac{L^2}{4}\rho_+\sqrt{\frac{\rho_+}{\rho_0}} = \frac{A_H}{4}.
\end{eqnarray}

\section{Charged Black Holes on ALF Gravitational Instantons and Their Thermodynamics}

In this section we construct a solution to the 5D Einstein-Maxwell-dilaton equations which represents a charged black hole on Taub-Bolt
instanton and examine its thermodynamics. A similar charged black hole on Taub-Nut instanton was already obtained and investigated in \cite{Yazadjiev:2006} using the same solution generation method.

\subsection{Solution generation}
The solution generation technique we will consider was originally developed in \cite{Yazadjiev:2005}. It allows the construction of static charged solutions of the EMd-equations from known solutions by exploiting the hidden symmetries of the dimensionally reduced action.

\paragraph{}The EMd gravity in $5$-dimensional spacetime is described by the action

\begin{equation}
I = -{1\over 16\pi} \int d^5x \sqrt{-g}\left(R - 2g^{\mu\nu}\partial_{\mu}\varphi \partial_{\nu}\varphi  -
e^{-2a\varphi}F^{\mu\nu}F_{\mu\nu} \right),
\end{equation}
and the following field equations can be derived

\begin{eqnarray} \label{FE}
R_{\mu\nu} &=& 2\partial_{\mu}\varphi \partial_{\nu}\varphi + 2e^{-2a\varphi} \left[F_{\mu\rho}F_{\nu}^{\rho} - {1\over 6}g_{\mu\nu} F_{\beta\rho} F^{\beta\rho}\right], \\
\nabla_{\mu}\nabla^{\mu}\varphi &=& -{a\over 2} e^{-2a\varphi} F_{\nu\rho}F^{\nu\rho}, \nonumber \\
&\nabla_{\mu}&\left[e^{-2a\varphi} F^{\mu\nu} \right]  = 0, \nonumber
\end{eqnarray}
where $R_{\mu\nu}$ is the Ricci tensor for the spacetime metric $g_{\mu\nu}$, $F_{\mu\nu}$ is the Maxwell tensor,
$\varphi$ is the dilaton field and $a$ is the dilaton coupling parameter.

We apply the solution generation technique \cite{Yazadjiev:2005} to a vacuum seed solution representing a static black hole on Taub-Bolt instanton (see section 5.1), and impose on the electromagnetic field the anzats

\begin{equation}\label{EF}
F = - d\Phi\wedge dt.
\end{equation}

In this way we obtain the following solution to the 5D EMd equations

\begin{eqnarray}
{ds}^{2}&=& - \left[1 + \frac{c(x-y)}{1+cx}\sinh^2\gamma \right]^{-{2\over 1 + a^2_{5}} }{\frac {  1+cy }{1+cx}}\,{dt}^{2}+
\left[1 + \frac{c(x-y)}{1+cx}\sinh^2\gamma \right]^{1\over 1 + a^2_{5} } \nonumber \\ \nonumber\\
&&\left[{\frac {F ( x,y ) }{H ( x,y ) }} \left( {d\psi}+\Omega d\phi \right) ^{2}+\frac{2{\kappa}^{4} \left( 1-c \right)  \left( 1+cx \right) H ( x,y ) }{ \left( 1-{\alpha}^{2} \right) \left( x-y \right) ^3} \left( {\frac {{dx}^{2}}{G ( x ) }}-{\frac {{dy}^{2}}{G ( y ) }}+A\,{{d\phi}}^{2} \right)\right]\nonumber,\\ \nonumber \\
e^{-2a\phi} &=& \left[1 + \frac{c(x-y)}{1+cx}\sinh^2\gamma \right]^{{2a_5^2\over 1 + a^2_{5}} }, \nonumber \\
\Phi &=& \frac{\sqrt{3}}{2}\frac{\sinh\gamma \cosh\gamma}{\sqrt{1+a_5^2}}\frac{c(x-y)}{1+cx}\left[1 + \frac{c(x-y)}{1+cx}\sinh^2\gamma \right]^{-1}
\end{eqnarray}

where the functions $G(x)$, $H(x,y)$, $F(x,y)$, $\Omega(x, y)$ and $A(x, y)$ are defined in the same way as in the vacuum case, as well as the ranges of the parameters and the C-metric like coordinates. The parameter $a_5$ is connected to the dilaton coupling parameter $a$ as $a_5 = \frac{\sqrt{3}}{2}a$. The physical infinity corresponds again to $(x, y)\rightarrow (-1, -1)$ and the length of the $S^1$ fibre at
infinity is $L = 8\pi n$.

The solution possesses the following asymptotic behavior in the coordinates we defined in ($\ref{asympt}$)
\begin{eqnarray}
ds^2 &=& - \left(1 - \frac{2\kappa^2 c}{r} - \frac{4\kappa^2 c}{r}\frac{\sinh^2\gamma}{1+a^2_5}\right)dt^2 + \nonumber \\
&+& \left( 1 - \frac{2m}{r} + \frac{\kappa^2 c}{r} + \frac{2\kappa^2 c}{r}\frac{\sinh^2\gamma}{1+a^2_5}\right)\left(d\psi + 2n\cos{\theta}d\phi\right)^2+ \nonumber \\
&+&\left(1 + \frac{2\kappa^2c}{r} + \frac{2\kappa^2 c}{r}\frac{\sinh^2\gamma}{1+a^2_5}\right)\left[\frac{dr^2}{\left( 1 - \frac{2m}{r} + \frac{\kappa^2 c}{r}\right)} + r^2 \left(d\theta^2 + \sin^2{\theta}d\phi^2\right)\right].
\end{eqnarray}

Using the asymptotical metric we compute the ADM mass and the tension by evaluating the Komar integrals ($\ref{MT}$). Thus we obtain

\begin{equation}\label{ADMC}
M_{ADM} = \frac{L}{2}\left( m + \frac{3}{2}c\kappa^2 + 3 c\kappa^2 \frac{\sinh^2\gamma}{1+a^2_5}\right)
\end{equation}
where the parameter $m$ is defined by ($\ref{parameter}$), while the tension coincides with the quantity calculated  for the vacuum case ($\ref{MA}$).
Again, the mass of the solution is calculated independently by the counterterm  method discussed in section 4.2. Direct calculation of the boundary
stress-energy tensor ($\ref{SET}$) leads to the result

\begin{equation}
8\pi T_t^t = \frac{1}{r^2}\left(m + \frac{3}{2}c \kappa^2 + 3c\kappa^2 \frac{\sinh^2\gamma}{1+a^2_5}\right) + O(\frac{1}{r^3}),
\end{equation}
hence

\begin{equation}
M_{ADM}= \frac{1}{8\pi}\int\left(m + \frac{3}{2}c \kappa^2 + 3c\kappa^2 \frac{\sinh^2\gamma}{1+a^2_5}\right)\sin{\theta}d\theta d\phi d\psi.
\end{equation}

Obviously performing the integration yields the same expression as obtained by the Komar approach.  In the limit $\gamma=0$ we recover the mass of the vacuum black hole on Taub-Bolt instanton, which was calculated in section 5.1. The components of the stress-energy tensor relevant for the calculation of the
tension coincide with the vacuum case ($\ref{SET1}$).

The intrinsic masses of the black hole and the instanton are obtained directly using ($\ref{MH}$) and ($\ref{MB}$) and found to coincide with the
 corresponding masses in the vacuum case

\begin{eqnarray}\label{MC}
M_H &=& M_H^0 = L c\kappa^2, \\
M_B &=& M_B^0 = {L\over 4}\kappa^2\left(1-c\right) \nonumber
\end{eqnarray}
The same feature is proved for the nut charge and the nut potential on the left and right semi-infinite intervals
obeying $N = n$ and $\chi_{L/R} = \chi^0_{L/R}$, where we denote by zero index the vacuum quantities.

\subsection{Smarr-like relation}

The derivation of the Smarr-like relation for charged black holes on an ALF gravitational instanton follows the same reasoning as in the vacuum case.
We obtain again the equation

\begin{eqnarray}
M_{ADM} =   {L\over 8} \int_{\hat{M}} \left[2di_\eta i_k \star d\xi
- d i_\eta i_\xi \star d k \right] - {L\over 8} \sum_i\int_{I_i}
\left[2i_\eta i_k \star d\xi - i_\eta i_\xi \star d k \right] ,
\end{eqnarray}
and interpret the boundary terms in the same way. The only difference is that the bulk term, which we proved to be

\begin{eqnarray}
{L\over 8} \int_{\hat{M}} \left[2di_\eta i_k \star d\xi - d i_\eta
i_\xi \star d k \right] =  {L\over 4} \int_{\hat{M}} \left[2i_\eta
i_k \star R[\xi] - i_\eta i_\xi \star R[k] \right],
\end{eqnarray}
does not vanish. Using the field equations we can obtain the following expression for the Ricci one-form

\begin{eqnarray}
\star R[\xi] = - 2e^{-2a\varphi} \left( -{2\over 3}i_{\xi}F\wedge \star F + {1\over 3} F\wedge i_{\xi}\star F \right).
\end{eqnarray}
Then, we take advantage of the fact that $i_kF =0$, $i_\xi\star F = 0$, and find
\begin{eqnarray}
i_k\star R[\xi]= -\frac{4}{3}e^{-2a\varphi} i_\xi F \wedge i_k\star F,
\end{eqnarray}

In a similar way we derive the corresponding expression for the Killing field $k$

\begin{eqnarray}
i_\xi\star R[k]= -\frac{2}{3}e^{-2a\varphi} i_\xi F \wedge i_k\star F,
\end{eqnarray}
and using the electromagnetic potential $\Phi$, defined by $d\Phi = - i_\xi F$ and $\Phi(\infty) = 0$, as well as the field equations $d\left(e^{-2a\varphi}i_{k}\star F\right)=0$ obtain

\begin{eqnarray}
&& {L\over 4} \int_{\hat{M}} \left[2i_\eta i_k \star R[\xi] - i_\eta i_\xi \star R[k] \right] =  - {L\over 2} \int_{\hat{M}}e^{-2a\varphi}i_\eta \left( i_\xi F \wedge i_k\star F \right)= \\
&=& {L\over 2} \int_{\hat{M}}e^{-2a\varphi}i_\eta \left( d \Phi\wedge i_k\star F \right) = - {L\over 2} \int_{\hat{M}}d \left( \Phi e^{-2a\varphi}i_\eta i_k\star F \right).
\end{eqnarray}

We can further simplify the last expression if we use the Stokes' theorem and consider that $i_\eta i_k \star F$ vanishes on the fixed point
sets of the spacelike Killing vectors, as well as that the integral over the 2-sphere at infinity tends to zero. Thus

\begin{eqnarray}
-{L\over 2} \int_{\hat{M}}d \left( \Phi e^{-2a\varphi}i_\eta i_k\star F \right) &=&
-{L\over 2} \int_{Arc(\infty)}  \Phi e^{-2a\varphi}i_\eta i_k\star F  -  {L\over 2}\sum_i \int_{I_i}  \Phi e^{-2a\varphi}i_\eta i_k\star F \nonumber\\
 &=& {L\over 2} \int_{I_H}  \Phi e^{-2a\varphi}i_k i_\eta\star F,
\end{eqnarray}

It can be proven that the electromagnetic potential is constant on the black hole horizon, so we finally obtain a simple expression for the bulk term

\begin{eqnarray}
 {L\over 2} \int_{I_H}  \Phi e^{-2a\varphi} i_k i_\eta \star F = {1 \over 4\pi} \Phi_H \int_{H} e^{-2a\varphi}\star F  = \Phi_H Q,
\end{eqnarray}
where we have defined the electric charge of the black hole as

\begin{eqnarray}\label{chargeg}
Q =  {1 \over 4\pi} \int_{H} e^{-2a\varphi}\star F.
\end{eqnarray}

Considering all the calculations we performed in the vacuum case the Smarr-like relations for the charged black hole on a Taub-Bolt instanton acquire the form

\begin{eqnarray}\label{SC}
M_{ADM} =  M_H + M_{B} + {L\over 2} N\chi + \Phi_H Q, \\
{\cal{T}}L =  {1 \over 2} M_H + 2M_{B} +  L N\chi. \nonumber
\end{eqnarray}

We can readily demonstrate the Smarr-like relation using the explicit quantities we calculated in the previous discussion of the solution.
If we evaluate the restriction of the electromagnetic potential on the horizon, we obtain, as expected, a constant value

\begin{eqnarray}
\Phi_H = \frac{\sqrt{3}}{2}\frac{\tanh{\gamma}}{{\sqrt{1+a_5^2}}}.
\end{eqnarray}

On the other hand, direct evaluation of ($\ref{chargeg}$) yields

\begin{eqnarray}\label{charge}
Q = \frac{\sqrt{3}\sinh\gamma\cosh\gamma}{\sqrt{1+a_5^2}} L \kappa^2c.
\end{eqnarray}

These results together with equations ($\ref{MC}$), and the expressions for the nut charge and potential verify the validity of the Smarr-like relations.
\subsection{Euclidean action}

The regularized action for 5D Einstein-Maxwell-dilaton gravity is given by

\begin{eqnarray}
I &=& - {1\over 16\pi} \int_M d^5x \sqrt{-g}\left(R - 2g^{\mu\nu}\partial_{\mu}\varphi \partial_{\nu}\varphi  -
e^{-2a\varphi}F^{\mu\nu}F_{\mu\nu} \right) - \nonumber \\
&-&{1\over 8\pi} \int_{\partial M} d^4x \sqrt{-h} \left(K - \sqrt{2{\cal R}}\right),
\end{eqnarray}
where $K$ is the external curvature on the boundary $\partial M$ and $\cal{R}$ is the Ricci scalar with respect to the boundary metric $h_{ij}$.
Direct calculations reveal that

\begin{eqnarray}
K &=& \frac{2}{r}\left(1 -\frac{1}{2}\frac{m}{r}-\frac{5}{4}\frac{\kappa^2 c}{r}  - \frac{3}{2}\frac{\sinh^2\gamma}{1 + a_5^2}\frac{c\kappa^2}{r}\right) \\
\cal{R} &=& \frac{2}{r^2}\left(1 - \frac{2\kappa^2 c}{r} - \frac{2\sinh^2\gamma}{1 + a_5^2}\frac{c\kappa^2}{r}\right),
\end{eqnarray}
and we obtain the boundary term $I_s$ of the action

\begin{eqnarray}
I_s &=&  - {1\over 8\pi} \int_{\partial M} d^4x \sqrt{-h} \left(K - \sqrt{2{\cal R}}\right) = \nonumber \\
&=& i\beta\frac{L}{2}\left(m + \frac{1}{2}\kappa^2c + \frac{\sinh^2\gamma}{1 + a_5^2}c\kappa^2\right),
\end{eqnarray}
where we have denoted with $\beta$ the period of the imaginary time coordinate equal to the inverse of the black hole temperature.

The bulk term $I_b$ can be expressed in a simple form if we take advantage of the field equations ($\ref{FE}$) and the symmetries of the electromagnetic tensor
\begin{eqnarray}
I_b &=& - {1\over 16\pi} \int_M d^5x \sqrt{-g}\left(R - 2g^{\mu\nu}\partial_{\mu}\varphi \partial_{\nu}\varphi  -
e^{-2a\varphi}F^{\mu\nu}F_{\mu\nu} \right) = \nonumber \\
&=& {1\over 4\pi} \int_M \frac{1}{3} e^{-2a\phi}F\wedge\star F = \frac{i\beta}{12\pi}\int_{M/U(1)}d\Phi\wedge e^{-2a\phi}\star F= \nonumber \\
&=& \frac{i\beta}{12\pi}\int_{\partial (M/U(1))} \Phi e^{-2a\phi}\star F = \frac{i\beta}{12\pi}\int_{S^3_\infty}\Phi e^{-2a\phi}\star F - \frac{i\beta}{12\pi}\int_{H} \Phi e^{-2a\phi} \star F = \nonumber \\
&=& - \frac{i\beta}{3}\Phi_HQ,
\end{eqnarray}
where we have considered  that the integral over the boundary at infinity tends to zero and the electromagnetic potential is constant on the horizon, and used the definition of the black hole charge.

Using the explicit expression for the restriction of the electromagnetic potential on the horizon and for the black hole charge ($\ref{charge}$) we obtain
\begin{equation}
I_b = - i\beta\frac{L}{2}\frac{\sinh^2\gamma}{1 + a_5^2}c\kappa^2
\end{equation}
leading to the final result for the Euclidean action
\begin{equation}
I = I_b + I_s = i\beta\frac{L}{2}\left(m + \frac{1}{2}\kappa^2c \right).
\end{equation}

The thermodynamical potential is determined by

\begin{equation}
{\cal{F}} = M - TS - \Phi_H Q = \frac{L}{2}\left(m + \frac{1}{2}\kappa^2c\right).
\end{equation}

and consequently

\begin{equation}\label{TSC}
TS  = \frac{L}{2}\kappa^2 c.
\end{equation}

The temperature of the black hole is computed by the surface gravity on the horizon according to

\begin{eqnarray}\label{TC}
T &=& \frac{\kappa_H}{2\pi} = \frac{1}{4\pi\kappa^2}\left(\cosh\gamma\right)^{-{3\over 1+ a_5^2}}\sqrt{\frac{1-\alpha^2}{2c(1+c)}}\nonumber \\
&=& \left(\cosh\gamma\right)^{-{3\over 1+ a_5^2}}T^0
\end{eqnarray}

and the black hole area is determined by straightforward integration as

\begin{eqnarray}\label{areac}
A_H &=& \int \sqrt{g_H}dx d\phi d\psi = \nonumber  \\
&=& 8\pi L \kappa^4 c \left(\cosh\gamma\right)^{{3\over 1+ a_5^2}}\sqrt{\frac{2c(1+c)}{1-\alpha^2}} = \left(\cosh\gamma\right)^{{3\over 1+ a_5^2}}A_H^0,
\end{eqnarray}
where we have denoted with zero index the corresponding quantities for the vacuum solution.

Comparing relations ($\ref{TSC}$), ($\ref{TC}$) and ($\ref{areac}$) we obtain that the entropy of the solution obeys

\begin{equation}
S = 2\pi L \kappa^4 c \left(\cosh\gamma\right)^{{3\over 1+ a_5^2}} \sqrt{\frac{2c(1+c)}{1-\alpha^2}} = \frac{A_H}{4}.
\end{equation}

\subsection{The First Law of Thermodynamics}

We will consider the so called 'physical process version' of the first law of thermodynamics developed by Wald \cite{Wald:1993}
which applies for a diffeomorphism covariant n-dimensional gravitational theory (see also \cite{Yazad:2009}). In such theory, one can associate  a Noether current $(n-1)$-form  and,
for solutions to the field equations, a Noether charge $(n-2)$-form  to every vector field $X$, both of which are locally constructed from $X$ and
the fields appearing in the Lagrangian ${\mathbf  L}$. The Noether current ${\cal I}^{X}$ can be expressed as

\begin{eqnarray}
{\cal I}^{X}= \Theta(\Gamma, {\cal L}_{X}\Gamma) - i_X {\mathbf L},
\end{eqnarray}
where the metric $g_{\mu\nu}$ and the matter fields are collectively denoted by $\Gamma$, and $d\Theta = \delta {\mathbf  L}$,
when the field equations are satisfied. If we choose $X$ to be a Killing field one can show that \cite{Wald:1993}

\begin{eqnarray}
\delta d{\cal N}^{X}=di_X\Theta.
\end{eqnarray}
which allows us to obtain the perturbations of the conserved quantities.

In our discussion we will restrict ourselves to the Lagrangian describing the solutions in the present article (see section 6)
\begin{eqnarray}
 {\mathbf  L} = \star R - 2d\varphi \wedge \star d\varphi - 2e^{-2a\varphi} F\wedge \star F,
\end{eqnarray}
where the electromagnetic field obeys the ansatz ($\ref{EF}$). It gives rise to the form $\Theta$

\begin{eqnarray}
d\Theta = d\star \,\upsilon - 4 (d\star d\varphi)\delta \varphi -
4\left(e^{-2a\varphi} \star F\right)\wedge  \delta F,
\end{eqnarray}
where
\begin{eqnarray}
\upsilon_\mu = \nabla^{\nu}\delta g_{\mu\nu} - g^{\alpha\beta}\nabla_{\mu}\delta g_{\alpha\beta}.
\end{eqnarray}
There are two relevant Noether forms for the calculation of the conserved quantities, ${\cal N}^\xi$ and ${\cal N}^k$ associated with the timelike
Killing field and the Killing field corresponding to the compact dimension respectively. They possess the form

\begin{eqnarray}
&&{\cal N}^{\xi}= - \star d\xi - 4\Phi \left(e^{-2a\varphi}\star F \right), \nonumber \\
&&{\cal N}^k= - \star dk,
\end{eqnarray}
where $\Phi$ is the electromagnetic potential defined as $d\Phi = - i_\xi F$ and $\Phi(\infty) = 0$.
It is convenient to use the 1-forms $i_\eta i_k{\cal N}^{\xi}$  and
$i_\eta i_\xi {\cal N}^k$ obeying

\begin{eqnarray}
\delta \left( di_\eta i_{k}{\cal N}^{\xi}\right)=di_\eta
i_{k}i_{\xi}\Theta, \;\;\; \; \; \delta \left(di_\eta i_{\xi}{\cal
N}^{k}\right)= - di_\eta i_{k}i_{\xi}\Theta,
\end{eqnarray}
and combine them to a single equality

\begin{eqnarray}\label{CI}
 \delta  \left(2 d i_\eta i_{k}{\cal N}^{\xi} - d i_\eta i_{\xi}{\cal N}^{k} \right)= 3di_\eta i_{k}i_{\xi}\Theta . \end{eqnarray}

Integrating on ${\hat M}$ and using the Stokes  theorem we obtain

\begin{eqnarray}\label{combinedidentity1}
\delta \int_{\partial {\hat M}} \left(2i_\eta i_{k}{\cal N}^{\xi} -
i_\eta i_{\xi}{\cal N}^{k} \right)= 3\int_{\partial {\hat M}} i_\eta i_{k}i_{\xi}\Theta.
\end{eqnarray}
Taking into account the explicit expressions of the Noether forms
we find

\begin{eqnarray}
&&\int_{Arc(\infty)} \left(2 i_\eta i_{k}{\cal N}^{\xi} - i_\eta i_{\xi}{\cal N}^{k} \right)= - 8 {M\over L} , \\
&&\int_{I_{H}} \left(2 i_\eta i_{k}{\cal N}^{\xi} - i_\eta i_{\xi}{\cal N}^{k} \right)=  8 {M_H\over L} + 16 \frac{\Phi_H}{L} Q , \\
&& \int_{I_{B}} \left(2 i_\eta i_{k}{\cal N}^{\xi} - i_\eta i_{\xi}{\cal N}^{k} \right)= 8 {M_B\over L}, \\
&& \int_{I_L\bigcup I_R} \left(2 i_\eta i_{k}{\cal N}^{\xi} - i_\eta
i_{\xi}{\cal N}^{k} \right)= 4 N\chi. \end{eqnarray} Respectively,
for  $i_k i_\xi \Theta$ we have

\begin{eqnarray}
&&\int_{Arc (\infty)}  i_\eta i_k i_\xi \Theta = 2(\delta c_t - \delta c_\psi) - 8 {\cal D} \delta\varphi_{\infty}, \\
&&\int_{I_{H}}  i_\eta i_k i_\xi \Theta =  \frac{1}{\pi}{{\cal A}_{H}\over L }\delta \kappa_{H} + 8 \frac{Q}{L}\delta\Phi_H, \\
&& \int_{I_{B}}i_\eta i_k i_\xi \Theta = {1\over \pi}{{\cal A}_{B} }
\delta \kappa_{B},
\end{eqnarray}
where $Q$ is the black hole charge, $\Phi_H$ is the restriction of the electromagnetic potential on the black hole horizon (see section 6.3), and $\varphi_{\infty}$ is the asymptotic value of the dilaton field. We have also taken into account that the dilaton charge can be expressed in the form

\begin{eqnarray}
{\cal D}= {1\over 4\pi} \int_{S^2_{\infty}} i_k i_\xi\star d\varphi.
\end{eqnarray}

We substitute these results in ($\ref{combinedidentity1}$) and take
into account that $3/4(\delta c_t - \delta c_\psi)=\delta (M/L) +
\delta {\cal T}$, where $c_t$ and $c_\psi$ are the coefficients in the first order asymptotic expansions of $g_{tt}$ and $g_{\psi\psi}$. Using the Smarr-like relations ($\ref{SC}$) and performing some algebraic manipulations we obtain the first law of thermodynamics for the mass and the tension

\begin{eqnarray}\label{1lawM}
\delta M &=&  {1\over 8\pi}\kappa_{H} \delta {\cal A}_{H}  + \Phi_H \delta Q   + L{\cal D}\delta\varphi_{\infty}, \\
\delta {\cal T}&=& -{1\over 8\pi}\frac{A_H}{L}\delta\kappa_{H} + {1\over 8\pi} \kappa_B \delta A_B  - \frac{Q}{L}\delta\Phi_H
 +  \frac{1}{2}N\delta\chi+ {\cal D}\delta\varphi_{\infty} \label{1lawT}.
\end{eqnarray}

In the derivation of the first laws (\ref{1lawM}) and (\ref{1lawT})
we have kept $L$ fixed.

\subsection{Extremal black hole on Taub-Bolt instanton}

The solution describing a charged black hole on a Taub-Bolt instanton allows an extremal case of black hole with zero temperature.
It is realized in the limit $\gamma \rightarrow \infty$,  $c\rightarrow0$ such as the parameter $c\sinh^2\gamma = q$ remains finite.
In this way we obtain the solution

\begin{eqnarray}
ds^2 &=& -\left[ 1 + \left(x-y\right)q \right]^{-\frac{2}{1+ a_5^2}}dt^2 + \left[ 1 + (x-y)q \right]^{\frac{1}{1+ a_5^2}}\nonumber \\ \nonumber \\
&&\left[{\frac {F (x,y ) }{H ( x,y ) }} \left( {d\psi}+\Omega d\phi \right) ^{2}\ + \frac{2{\kappa}^{4} H ( x,y ) }{ \left( 1-{\alpha}^{2} \right) \left( x-y \right) ^3} \left( {\frac {{dx}^{2}}{G ( x ) }}-{\frac {{dy}^{2}}{G ( y ) }}+A\,{{d\phi}}^{2} \right)\right], \nonumber\\
\Phi &=& \frac{\sqrt{3}}{2\sqrt{1+a_5^2}}\frac{(x-y)q}{\left(1+ (x-y)q\right)}, \nonumber \\
e^{-2a\phi} &=& \left[ 1 + (x-y)q\right]^{{2a_5^2\over 1 + a^2_{5}} },
\end{eqnarray}
where the metric functions have the following form
\begin{eqnarray}
G ( x ) &=& ( 1-{x}^{2} )\,,  \\
H ( x,y ) &=&  \left( 1-y \right)^{2}-{
\alpha}^{2}   \left( 1-x \right)^{2}, \nonumber \\
F ( x,y ) &=& ( 1-{\alpha}^{2} )  \left( 1-x
 \right)  \left( 1-y \right), \nonumber \\
\Omega&=&{\frac {2 \alpha\,{\kappa}^{2} \left(2+x+y\right)}{ \left( 1-{\alpha}^{2}
 \right)  \left( x-y \right) }} , \nonumber \\
A&=&-{\frac {2 \left( 1+x \right)  \left( 1+y \right) }{
 \left( x-y \right) }}. \nonumber
\end{eqnarray}

The coordinates take values in the ranges $-1 < x < 1$, $-\infty < y < -1$,  physical infinity corresponds to $(x, y) = (-1,-1)$, and
the black hole horizon is located at $ y \rightarrow -\infty$. The dilaton field is however divergent at $y\rightarrow -\infty$, if the parameter $a_5$ does not vanish. Therefore, the extremal black hole exists only within Einstein-Maxwell gravity corresponding to $a_5 = 0$.

We will present the extremal solution in the more convenient coordinates $r$ and $\theta$ defined by ($\ref{parameter}$) as

\begin{eqnarray}
ds^2 &=& - \left[ 1 + h(r, \theta)q\right]^{-2}dt^2 + \nonumber \\
&+&\left[ 1 + h(r,\theta)q\right]\left[f(r)\left(d\psi + 2n\cos{\theta}d\phi\right)^2+ \frac{1}{f(r)}dr^2 + \left(r^2 - n^2 \right) \left(d\theta^2 + \sin^2{\theta}d\phi^2\right)\right], \nonumber \\
\Phi &=& \frac{\sqrt{3}}{2}\frac{h(r, \theta)q}{1+ h(r, \theta)q},
\end{eqnarray}

where

\begin{eqnarray}
f(r) = \frac{\left(r^2 + n^2- 2mr\right)}{\left(r^2 - n^2\right)},  \\
h(r, \theta) = \frac{2(1-\alpha^2)}{\frac{\alpha}{n}r -1 + (1-\alpha^2)\cos^2{\theta\over 2}}. \nonumber
\end{eqnarray}

The horizon is located at $r = \frac{n}{\alpha}$, $\theta = \pi$. We introduce the near-horizon coordinates

\begin{eqnarray}
r =\frac{n}{\alpha} + 2\kappa^2 y_1^2, \quad \theta = \pi + 2y_2,
\end{eqnarray}
where $y_1$ and $y_2$ are small, and redefine the coordinates once again as

\begin{eqnarray}
R = y_1^2 + \frac{1}{2}y_2^2, \quad
\theta' = \arctan{\left(\sqrt{\frac{1}{2}}\frac{y_2}{y_1}\right)}, \quad
\psi' = \frac{(1-\alpha^2)}{2\kappa^2}\left(\psi - 2n\phi\right).
\end{eqnarray}

Performing a further coordinate transformation

\begin{equation}
dt = d\tau - \sqrt{\frac{2\kappa^4q^3}{(1-\alpha^2)}}d\left(\frac{1}{R}\right)
\end{equation}

the metric acquires the form

\begin{eqnarray}
ds^2 = - \frac{R^2}{q^2}d\tau^2 -2\sqrt{\frac{2\kappa^4}{q(1-\alpha^2)}}d\tau d R + \frac{8\kappa^4q}{(1-\alpha^2)}\left(d\theta'^2 + \sin^2\theta'd\phi^2 + \cos^2\theta'd\psi'^2\right),
\end{eqnarray}
which is clearly regular. For regular solutions corresponding to $\alpha = {1\over 2}$ (see section 5.1) the angular variable $\theta$ takes the range $0 \leq \theta' \leq \pi/2$, and $\Delta \phi =  2\pi$, $\Delta  \psi' = 2\pi$. Therefore, an event horizon with $S^3$ topology is located at $R = 0$.
We can calculate the physical characteristics of the solution by taking the extremal limit in the corresponding expressions.
Thus we obtain for the charge and the horizon area

\begin{eqnarray}
|Q| &=& \sqrt{3}L \kappa^2q, \nonumber \\
A_H &=& 4\sqrt{2\pi} \left(\frac{|Q|}{\sqrt{3}}\right)^{{3\over 2}},
\end{eqnarray}
using the regularity condition $\alpha = \frac{1}{2}$. The instanton mass and ADM mass reduce to

\begin{eqnarray}
 M_B &=& \frac{L}{4} \kappa^2, \nonumber \\
M_{ADM} &=& \frac{L}{2} m + \frac{\sqrt{3}}{2}|Q|.
\end{eqnarray}
The temperature of the extremal black hole vanishes as well as its Komar mass.

\subsection{Extremal black hole on Taub-Nut instanton}

Although the solution describing a charged black hole on a Taub-Nut instanton was generated in \cite{Yazadjiev:2006}, its extremal limit was never investigated.
Actually, an extremal solution exists and it is realized in similar way as in the case of Taub-Bolt instanton.

The metric of the charged black hole on a Taub-Nut instanton has the form \cite{Yazadjiev:2006}

\begin{eqnarray}
ds^2 &=& - {\left(1 - {\rho_{+}\over \rho}\right)\over \left[1 + {\rho_{+}\over \rho}\sinh^2\gamma\right]^{2\over 1+ a^2_{5}}} dt^2
 + \left[1 + {\rho_{+}\over \rho}\sinh^2\gamma\right]^{1\over 1+ a^2_{5}} \nonumber \\
&&\left[{ \left(1 + {\rho_{0}\over \rho }\right)\over \left(1 - {\rho_{+}\over \rho }\right)} d\rho^2 + \rho (\rho + \rho_{0})\left( d\theta^2 + \sin^2\theta d\phi^2\right) +
{\rho\over \rho + \rho_{0}}\left( d\psi + r_{\infty} \cos\theta d\phi\right)^2 \right] ,\nonumber \\ \nonumber \\
\Phi &=& {\sqrt{3}\cosh\gamma\sinh\gamma\over 2\sqrt{1 + a^2_{5}}} {\rho_{+}\over \rho + \rho_{+}\sinh^2\gamma }, \nonumber \\
e^{-2a\phi} &=& \left[1 + {\rho_{+}\over \rho}\sinh^2\gamma \right]^{{2a_5^2} \over 1 + a^2_{5}},
\end{eqnarray}
where $r_{\infty} = \sqrt{\rho_0 (\rho_0 + \rho_{+})}$ and the black hole horizon is located at $\rho = \rho_+$, $0\leq \theta \leq \pi $.
The extremal limit is realized when $\gamma \rightarrow \infty$ and  $\rho_+\rightarrow0$ in such a way that the parameter $\rho_+\sinh^2\gamma = q$ remains finite. Thus the extremal solution becomes

\begin{eqnarray}
ds^2 &=& -  \left[1 + {q\over \rho}\right]^{-{2\over 1+ a^2_5}} dt^2
 + \left[1 + {q\over \rho}\right]^{1\over 1+ a^2_{5}} \nonumber \\
&&\left[ \left(1 + {\rho_{0}\over \rho }\right) d\rho^2 + \rho (\rho + \rho_{0})\left( d\theta^2 + \sin^2\theta\phi^2\right) +
 {\rho\over \rho + \rho_{0}}\left( d\psi + r_{\infty}\cos\theta d\phi\right)^2 \right] ,\nonumber \\
\Phi &=& {\sqrt{3}\over 2\sqrt{1 + a^2_{5}}} {q\over \rho + q }, \nonumber \\
e^{-2a\phi} &=& \left[1 + {q\over \rho} \right]^{2a^2_{5} \over 1 + a^2_{5}}.
\end{eqnarray}

The radial coordinate takes the range $0\leq \rho < \infty$ and the black hole horizon is located at $\rho = 0$. The extremal black hole exists only in the Einstein-Maxwell case since the dilation field diverges at $a_5 \neq 0$. It is easy to prove that the metric is regular at $\rho = 0$
by expanding it near the horizon. Using the coordinates $\rho = R$, $R\ll 1$ and $\psi' = \psi/ r_\infty$, and the transformation $dt = d\tau - \sqrt{\rho_0 q^3}d\left(\frac{1}{R}\right)$ the metric acquires the form

\begin{eqnarray}
ds^2 = -  \frac{R^2}{q^2} d\tau^2 -2\sqrt{\frac{\rho_0}{q}}dR d\tau + q\rho_0\left[ d\theta^2 + \sin^2\theta d\phi^2 + \left(d\psi' + \cos\theta d\phi\right)^2 \right],
\end{eqnarray}
where the angular variable $\theta$ takes the range  $0\leq \theta \leq \pi$, and $\Delta\phi = 2\pi$, $\Delta \psi' = 4\pi$. The metric remains regular at $R\rightarrow 0$, consequently a horizon with $S^3$ topology is located at $R = 0$.
The physical characteristics obtained by taking the extremal limit are

\begin{eqnarray}
|Q| &=& 2\pi\sqrt{3}q\rho_0, \nonumber\\
A_H &=& 4\sqrt{2\pi}\left(\frac{|Q|}{\sqrt{3}}\right)^{3\over 2}, \nonumber \\
M_{ADM} &=& \pi\rho_0^2 + \frac{\sqrt{3}}{2} |Q|.
\end{eqnarray}
The temperature of the extremal black hole vanishes as well as its Komar mass.

\section{Conclusion}

In the present article we have examined some of the thermodynamical properties of solutions to the 5D Einstein and Einstein-Maxwell-dilaton equations describing a static black hole on a ALF gravitational instanton. We argued that the nut charge and the nut potential are important physical characteristics of these solutions and developed techniques for their calculation. Using the Komar definition of the mass and the tension we derived a Smarr-like relation and demonstrated that the nut charge and potential contribute to it. Two particular solutions describing static black holes on Taub-Nut and Taub-Bolt instantons are investigated as examples. The investigation includes computation of the conserved charges  in two independent ways - by Komar and counterterm approach and  comparison of the results. The Euclidean action is also obtained and used to determine the entropy of the solutions. Finally, a new solution to the 5D EMd equations representing a charged black hole on a Taub-Bolt instanton was generated and examined.
It was shown that within the Einstein-Maxwell gravity it possesses a well-defined extremal limit.

\section*{Acknowledgements}
 The partial support by the Bulgarian National Science Fund under Grant DO 02-257 and Sofia University Research Fund under Grant No 88/2011,
 is gratefully  acknowledged.

\end{document}